\documentclass[12pt]{article}
\begin{document}

\title{\bf The Pseudo-Newtonian Force and Potential
about a Higher Dimensional Rotating Black Hole}

\author{M. Sharif \thanks{e-mail: msharif@math.pu.edu.pk}
\\ Department of Mathematics, University of the Punjab,\\
Quaid-e-Azam Campus Lahore-54590, PAKISTAN.}

\date{}

\maketitle
\begin{abstract}
In this paper, we study the behavior of the pseudo-Newtonian force
and potential about a higher dimensional rotating black hole. We
obtain conditions for the force character from an attractive to
repulsive. We also find the conditions under which force attains a
maximum value. The results of this paper generalizes the already
found structure of force and potential about a five dimensional
rotating black hole. It is interesting to note that we recover the
five dimensional results under a special case.
\end{abstract}


\section{Introduction}

The idea of re-introducing the Newtonian gravitational force into
the theory of General Relativity (GR) arose in an attempt to deal
with the following problem: Gravitation, being non-linear, should
dominate over the Coulomb interaction at some, sufficiently small,
scale. At what scale would it occur? Whereas this question is
perfectly valid in pre-relativistic terms it becomes meaningless in
GR. The reason is that gravitation is expressed in purely geometric
terms [1] while electromagnetism is not. Thus, in Relativity,
gravitation possesses a very different status than the other forces
of Nature. Our physical intuition for the other interactions,
nevertheless, rests on the concept of forces. To deal with gravity
and other forces together, we must either express the other forces
geometrically, as in the Kaluza-Klein theories, or express
gravitation in the same terms as the other forces. We will follow
the latter alternative as the simpler program to implement.

The procedure adopted converts an idealised operational definition
of the gravitational force (via the tidal force) into a mathematical
formulation. In the pseudo-Newtonian ($\psi N$) approach [2,3], the
curvature of the spacetime is {\it straightened out} to yield a
relativistic force which bends the path, so as to again supply the
guidance of the earlier, force-based, intuition. Some insights have
already been obtained [2-5] by expressing the consequences of GR in
terms of forces by applying it to Kerr and Kerr-Newmann metrics. The
$\psi N$ potential has also been evaluated for charged particle in
Kerr-Newmann geometry by Ivanov and Prodanov [6]. In a recent paper
[7], the structure of the $\psi N$ force and potential has been
analyzed about a five dimensional rotating black hole and some
insights have been achieved. This paper investigates the structure
of the $\psi N$ force and potential about a higher dimensional
rotating black hole.

The paper is organized as follows: In the next section, we briefly
discuss the $\psi N$ formalism so that the definition of force and
potential can be given in this formalism. Section $3$ provides the
details of the metric describing higher dimensional rotating black
hole. We calculate the force and potential for this metric in
section $4$. Finally, section $5$ concludes the results.

\section{The Pseudo-Newtonain Formalism}

The basis of the formalism is the observation that the tidal force,
which is operationally determinable, can be related to the curvature
tensor by
\begin{equation}
F_T^\mu=mR_{\nu\rho\pi}^\mu t^\nu l^\rho t^\pi,\quad
(\mu,\nu,\rho,\pi=0,1,2,3),
\end{equation}
where $m$ is the mass of a test particle, $t^\mu=f(x)\delta_0^\mu,~
f(x)=(g_{00})^{-1/2}$ and $l^\mu$ is the separation vector. $l^\mu$
can be determined by the requirement that the tidal force have
maximum magnitude in the direction of the separation vector.
Choosing a gauge in which $g_{0i}=0$ (similar to the synchronous
coordinate system [8]) in a coordinate basis. We further use Riemann
normal coordinates (RNCs) for the spatial direction, but not for the
temporal direction. The reason for this difference is that both ends
of the accelerometer are spatially free, i.e., both move and do not
stay attached to any spatial point. However, there is a
\emph{memory} of the initial time built into the accelerometer in
that the zero position is fixed then. Any change is registered that
way. Thus \emph{time} behaves very differently from \emph{space}.

The relativistic analogue of the Newtonian gravitational force
called the $\psi N$ gravitational force, is defined as the quantity
whose directional derivative along the accelerometer, placed along
the principal direction, gives the extremised tidal force and which
is zero in the Minkowski space. Thus the $\psi N$ force, $F^\mu$,
satisfies the equation
\begin{equation}
F_T^{*\mu}=l^\nu F_{;\nu}^\mu ,
\end{equation}
where $F_T^{*\mu}$ is the extremal tidal force corresponding to the
maximum magnitude reading on the dial. Notice that $F_T^{*0}=0$ does
not imply that $F^0=0$. With the appropriate gauge choice and using
RNCs spatially, Eq.(2) can be written in a space and time break up
as
\begin{eqnarray}
l^i(F_{,i}^0+\Gamma_{ij}^0F^j)&=&0,\\
l^j(F_{,j}^i+\Gamma_{0j}^iF^0)&=&F_T^{*i},\quad (i,j=1,2,3).
\end{eqnarray}
A simultaneous solution of the above equations can be found by
taking the ansatz [3]
\begin{eqnarray}
F_0&=&-m\left[\{\ln
(Af)\}_{,0}+g^{ik}g_{jk,0}g^{jl}g_{il,0}/4A\right],\\
F_i&=&m(\ln f)_{,i},
\end{eqnarray}
where $A=(\ln \sqrt{-g})_{_{,0}},~ g=det(g_{_{ij}})$. It is
mentioned here that this force formula depends on the choice of
frame, which is not uniquely fixed.

The spatial component of the $\psi N$ force $F_i$ is the
generalisation of the force which gives the usual Newtonian force
for the Schwarzschild metric and a $\frac{Q^2}{r^3}$ correction to
it in the Riessner-Nordstrom metric. The $\psi N$ force may be
regarded as the \emph{Newtonian fiction} which \emph{explains} the
same motion (geodesic) as the \emph{Einsteinian reality} of the
curved spacetime does. We can, thus, translate back to Newtonian
terms and concepts where our intuition may be able to lead us to
ask, and answer, questions that may not have occurred to us in
relativistic terms. Notice that $F_i$ does not mean deviation from
geodesic motion.

The quantity whose proper time derivative is $F_\mu$ gives the
momentum four-vector for the test particle. Thus the momentum
four-vector, $p_\mu$, is [4]
\begin{equation}
p_\mu=\int F_\mu dt.
\end{equation}
The zero component of the momentum four-vector corresponds to the
energy imparted to a test particle of mass $m$ while the spatial
components of this vector give the momentum imparted to test
particles as defined in the preferred frame (in which
$g_{_{0i}}=0)$. We can also write Eqs.(5) and (6) as follows
\begin{equation}
F_0=-U_{,0},\quad F_i=-V_{,i},
\end{equation}
where the quantities $U$ and $V$ are given by
\begin{eqnarray}
U&=&m[ln
(Af/B)+\int(g^{ij}_{,0}g_{ij,0}/4A)dt],\\
V&=&-m\ln f.
\end{eqnarray}
Here $B$ is a constant with units of time inverse so as to make
$A/B$ dimensionless.

In the free fall rest-frame, the $\psi N$ force is given [3,4] by
\begin{equation}
F_i=-m(\ln \sqrt {g_{00}})_{,i}=-V_{,i},
\end{equation}
where
\begin{equation}
V=m(\ln \sqrt {g_{00}}).
\end{equation}
This quantity $V$ gives the generalization of the classical
gravitational potential and, for small variations from Minkowski
space
\begin{equation}
V\approx\frac{1}{2}m(g_{00}-1)
\end{equation}
which is the pseudo-Newtonian potential. We shall analyse the
behavior of these quantities for the higher dimensional rotating
black hole in the next section.

\section{Higher Dimensional Rotating Black Hole}

The metric of a rotating black hole in higher dimensions follows
from the general asymptotically flat solutions to $(N+1)$
dimensional vacuum gravity found by Myers and Perry [9]. We consider
higher dimensional rotating black hole with a single rotation
parameter. In Boyer-Lindquist type coordinates, the metric is given
by [10]
\begin{eqnarray}
ds^2&=&(1-\frac{M}{r^{N-4}\Sigma})dt^2-\frac{r^{N-2}\Sigma}{\Delta}dr^2-\Sigma
d\theta^2\nonumber\\
&-&(r^2+a^2+\frac{Ma^2\sin^2\theta}{r^{N-4}\Sigma})\sin^2
\theta d\phi^2\nonumber\\
&+&\frac{2Ma\sin^2\theta}{r^{N-4}\Sigma}dtd\phi-r^2\cos^2\theta
d\Omega^2_{N-3},
\end{eqnarray}
where
\begin{equation}
\Sigma=r^2+a^2\cos^2\theta, \quad \Delta=r^{N-2}(r^2+a^2)-Mr^2,
\end{equation}
and $M$ is a parameter related to the physical mass of the black
hole, while the parameter $a$ is associated with its angular
momentum. The quantity
\begin{equation}
d\Omega^2_{N-3}=d\chi_1^2+\sin^2\chi_1(d\chi_2^2+\sin^2\chi_2(...d\chi_{N-3}^2...))
\end{equation}
represents the metric of a unit $(N-3)$-sphere.

It is mentioned here that for $N=3$, neglecting the last term of
Eq.(14), this gives the analogue of the Kerr metric. Also, for
$N=4$, this exactly reduces to the metric representing five
dimensional rotating black hole with $b=0$ [7]. The event horizon of
the black hole is a null surface determined by the equation
\begin{equation}
\Delta=r^2+a^2-\frac{M}{r^{N-4}}=0.
\end{equation}
The largest root of this equation gives the radius of the black
hole's outer event horizon. Notice that for $N=3$ and $N=4$, the
horizon exists unless its rotation achieves the maximum speed by the
mass of the black hole. For $N\geq 5$, the horizon exists
independent of the rotation [9,11]. The rotational symmetry in the
$\phi$-direction along with the time-translation invariance of the
metric imply the existence of the commuting Killing vectors
\begin{equation}
\xi_{(0)}=\xi^\mu_{(t)}\frac{\partial}{\partial x^\mu}, \quad
\xi_{(3)}=\xi^\mu_{(\phi)}\frac{\partial}{\partial x^\mu}.
\end{equation}
The scalar products of these Killing vectors can be written down in
terms of the metric components as given below:
\begin{eqnarray}
\xi_{(0)}.\xi_{(0)}&=&g_{00}=1-\frac{M}{r^{N-4}\Sigma}, \nonumber\\
\xi_{(0)}.\xi_{(3)}&=&g_{03}=\frac{Ma\sin^2\theta}{r^{N-4}\Sigma}, \nonumber\\
\xi_{(3)}.\xi_{(3)}&=&g_{33}=-(r^2+a^2+\frac{Ma^2\sin^2\theta}{r^{N-4}\Sigma})\sin^2\theta.
\end{eqnarray}
The Killing vectors (18) can be used to give a physical
interpretation of the parameters $M$ and $a$ involved in the metric
(14). The following coordinate-independent definitions for these
parameters can be obtained by using the analysis given in the paper
[12]. Thus
\begin{equation}
M=\frac{1}{(N-2)A_{N-1}}\oint\xi^{\mu;\nu}_{(t)}d^{N-1}\Sigma_{\mu\nu}
\end{equation}
and
\begin{equation}
j=aM=-\frac{1}{4\pi^2}\oint\xi^{\mu;\nu}_{(\phi)}d^{N-1}\Sigma_{\mu\nu}.
\end{equation}
The integrals are taken over the $(N-1)$-sphere at spatial infinity
\begin{equation}
d^{N-1}\Sigma_{\mu\nu}=\frac{1}{(N-1)!}\sqrt{-g} \epsilon_{\mu\nu
i_1i_2...i_{N-1}}dx^{i_1}\wedge dx^{i_2}\wedge...\wedge\wedge
dx^{i_{N-1}}
\end{equation}
and
\begin{equation}
A_{N-1}=\frac{2\pi^{N/2}}{\Gamma(N/2)}
\end{equation}
is the area of a unit $(N-1)$-sphere. These definitions can be
verified by evaluating the integrands in the asymptotic region
$r\rightarrow \infty$. The dominant terms in the asymptotic
expansion take the following form
\begin{eqnarray}
\xi_{(t)}^{t;r}&=&\frac{M(N-2)}{2r^{N-1}}+O(\frac{1}{r^{N+1}}), \nonumber\\
\xi_{(\phi)}^{t;r}&=&-\frac{jN\sin^2\theta}{2r^{N-1}}+O(\frac{1}{r^{N+1}}).
\end{eqnarray}
It is obvious that these expressions justify Eqs.(20) and (21). The
total mass $M_T$ and the total angular momentum $J$ of the black
hole can be found [9] as
\begin{equation}
M=\frac{16\pi GM_T}{(N-1)A_{N-1}},\quad j=\frac{8\pi GJ}{A_{N-1}}.
\end{equation}
This equation justifies the interpretation of the parameters $M$ and
$a$ related to the physical mass and angular momentum of the black
hole.

\section{The Pseudo-Newtonain Force and Potential}

In this section, we calculate $\psi N$ force and potential for the
higher dimensional rotating black hole and analyze them. Using
Eqs.(11) and (14), the structure of the $\psi N$ force (per unit
mass of the test particle) for the higher dimensional rotating black
hole takes the following form
\begin{equation}
F_r=-\frac{M[(N-4)r^{N-3}\Sigma+2r^{N-3}]}{2r^{N-4}\Sigma(r^{N-4}\Sigma-M)},
\end{equation}
\begin{equation}
F_\theta=\frac{Ma^2\sin\theta\cos\theta}{r^{N-4}\Sigma(r^{N-4}\Sigma-M)}.
\end{equation}
It follows from Eq.(26) that the radial component can never become
zero outside the horizon and hence this force cannot change
character from an attractive to a repulsive outside the black hole.
From Eq.(27), we see that the polar component can only become zero
outside the horizon at $\theta=0,~\pi/2,~\pi$. Notice that naked
singularities can give repulsive as well as attractive forces. When
the turnover lies outside the horizon, the structure of force can
provide interesting features for its optimal values according to $r$
or $\theta$. Since our observers are seeing {\it force} in a flat
space, the metric to be used is the plane polar one. The square of
the magnitude of the force is
\begin{equation}
(F)^2=\frac{M^2}{4r^{2N-6}\Sigma^2(r^{N-4}\Sigma-M)^2}[r^{2N-4}((N-4)\Sigma
+2)^2 +4a^4\sin^2\theta\cos^2\theta]
\end{equation}
which implies that
\begin{equation}
|F|=\frac{M}{2r^{N-3}\Sigma(r^{N-4}\Sigma-M)}[r^{2N-4}((N-4)\Sigma
+2)^2 +4a^4\sin^2\theta\cos^2\theta]^{1/2}.
\end{equation}
When we expand it in powers of $(1/r)$, it follows that
\begin{equation}
|F|=\frac{Mr((N-4)\Sigma+2)}{2\Sigma(r^{N-4}\Sigma-M)}[1+\frac{2a^4}{r^{2N-4}((N-4)\Sigma
+2)^2}\sin^2\theta\cos^2\theta+....].
\end{equation}

The equations for the turnovers along $r$ and $\theta$,
respectively, are
\begin{eqnarray}
&&r^{2N-4}\Sigma(r^{N-4}\Sigma-M)((N-4)\Sigma+2)[(2N-4)((N-4)\Sigma+2)\nonumber\\
&+&4(N-4)]-[r^{2N-4}((N-4)\Sigma+2)^2+4a^4\sin^2\theta\cos^2\theta]\nonumber\\
&\times&[(2N-6)\Sigma(r^{N-4}\Sigma-M)
+4r^2(r^{N-4}\Sigma-M)\nonumber\\
&+&2r^{N-2}\Sigma((N-4)\Sigma+2)]=0,
\end{eqnarray}
\begin{eqnarray}
&&2\Sigma(r^{N-4}\Sigma-M)[a^2\cos2\theta-r^{2N-4}(N-4)((N-4)\Sigma+2)]\nonumber\\
&+&(r^{N-4}\Sigma-M+\Sigma)[4a^4\sin^2\theta\cos^2\theta+r^{2N-4}((N-4)\Sigma+2)^2]=0.
\end{eqnarray}
The complexity of Eqs.(31) and (32) makes it difficult to analyse
them generally. However, we can investigate these equations for the
following two special cases.\\
\par \noindent
(i)\quad$N=4,\quad a\neq0$,\quad (ii)\quad$N=4,~a=0$.\\
\par \noindent
The first case (i) exactly reduces to the results of the paper [7].
In this case, it follows from Eqs.(31) and (32) that
\begin{equation}
(r^2+a^2)(r^2+a^2-M)-2r^2(2r^2+2a^2-M)=0,
\end{equation}
\begin{equation}
2(r^2+a^2)-M=0.
\end{equation}
The first equation is satisfied when
\begin{equation}
r^2=\frac{1}{6}[(M-2a^2)\pm\sqrt{M^2+16a^4-16Ma^2}].
\end{equation}
Eq.(34) is satisfied for the value of $r$ given by
\begin{equation}
r^2=\frac{1}{2}M-a^2.
\end{equation}
For $N=4,~a=0$, i.e., when there is no rotation, Eqs.(31) and (32),
respectively, are satisfied for the values of $r$ given as
\begin{equation}
r=\pm\sqrt{\frac{M}{3}},\quad r=\pm\sqrt{\frac{M}{2}}.
\end{equation}
In the first case (i), a maximum of the magnitude can be achieved at
the value of $r$ given by Eq.(35) and hence the maximum value of the
force can be obtained by replacing this value of $r$ in $F_r$. For
the case (ii), a maximum value of the magnitude occurs at $r$ as
given in Eq.(37).

The corresponding potential is obtained by using Eqs.(13) and (14)
and is given by
\begin{equation}
V=-\frac{M}{2r^{N-4}\Sigma}.
\end{equation}
For the special case (i), it becomes
\begin{equation}
V=-\frac{M}{2(r^2+a^2)}.
\end{equation}
When there is no rotation, it reduces to
\begin{equation}
V=-\frac{M}{2r^2}.
\end{equation}
It is remarked that the structure of force and potential indicate
similar type of behavior as for the Kerr metric [4] in the special
cases.

\section{Summary}

The relativistic analogue of the Newtonian gravitational potential
is related to the square of the magnitude of the timelike Killing
vector [13]. This conjecture is verified as an approximation and has
been applied to the Schwarzschild metric. Later, its validity was
justified for a particular class of spacetimes [5]. Thus the
procedure of looking for the implications of relativity in terms of
an analogue of the Newtonian gravitational force [4] has already
provided some insights [1-5,14].

This paper is particularly emphasized to analyse the structure of
the pseudo-Newtonian force and potential about a higher dimensional
rotating black hole. When we take $N=4$, this gives the same results
as given in the paper [6] with only one angular momentum. We have
found that the radial component of the force cannot change character
from an attractive to a repulsive outside the black hole as it never
becomes zero there. The polar component becomes zero outside the
horizon at $\theta=0,~\pi/2,~\pi$. When the turnover lies outside
the horizon, the structure of the force can provide interesting
features for its optimal values according to $r$ or $\theta$. The
equations for the turnovers along $r$ and $\theta$ are too difficult
to analyse. However, we have investigated these equations for the
two special cases, i.e., $N=4,\quad a\neq0$ and $N=4,~a=0$. The
first case exactly reduces to the results of the paper [7] while the
second case corresponds to no rotation. It is worth mentioning that
the general case gives similar type of behavior of the force and
potential as for the Kerr metric [5] in the special cases.



\vspace{2cm}

{\bf \large References}

\begin{description}

\item{[1]} Misner, C.W., Thorne, K.S. and Wheeler, J.A.: {\it Gravitation}
(W.H. Freeman, 1973).

\item{[2]} Qadir, A. and Quamar, J.: {\it Proc. 3rd Marcel Grossmann Meeting on
General Relativity}, ed. Hu Ning (North Holland Amstderm 1983)189;\\
Quamar, J.: {\it Ph.D. Thesis} Quaid-i-Azam University Islamabad
(1984).

\item{[3]} Qadir, A. and Sharif, M.: Nuovo Cimento {\bf B107}(1992)1071;\\
Sharif, M.: {\it Ph.D. Thesis} Quaid-i-Azam University Islamabad
(1991).

\item{[4]} Qadir, A. and Sharif, M.: Phys. Lett. {\bf A167}(1992)331;\\
Sharif, M.: Astrophys. and Space Science {\bf 253}(1997)195.

\item{[5]} Qadir, A. and Quamar, J.: Europhys. Lett. {\bf
2}(1986)423;\\
Qadir, A.: Europhys. Lett. {\bf 2}(1986)427.

\item{[6]} Ivanov, Rossen I. and Prodanov, M.: Phys. Lett. {\bf B611}(2005)34.

\item{[7]} Sharif, M.: Nuovo Cimento {\bf B121}(2006)121.

\item{[8]} Landau, L.D. and Lifschitz, E.M. {\it
The Classical Theory of Fields} (Pergamon Press 1975).

\item{[9]} Myers, R.C. and Perry, M.J.: Ann. Phys. {\bf172}(1986)304.

\item{[10]} Aliev, A.N.: Phys. Rev. \textbf{D74}(2006)024011.

\item{[11]} Emparan, R. and Myers, R.C.: J. High Energy Phys. {\bf 0309}(2003)25.

\item{[12]} Komar, A.: Phys. Rev. {\bf 113}(1959)934.

\item{[13]} Hawking, S.W. and Ellis, G.F.R.: {\it The Large Scale Structure of Spacetime}
(Cambridge University Press, 1973).

\item{[14]} Mahajan, S.M., Qadir, A. and Valanju, P.M.: Nuovo Cimento {\bf
B65}(1981)404.
\end{description}

\end{document}